# Strategic and Operational information support of decision making processes and systems

**Omar Abahmane**
ISIAM, Av. Hassan I, B.P 805, Agadir, Maroc
laraid@isiam.ma
Institut Supérieur d'Informatique Appliquée et de Management ISIAM, Agadir, Maroc
FSA, Université Ibn Zohr Agadir, Maroc
Laboratoire de Recherché Appliquée en Informatique Décisionnelle (LARAID) Agadir, Maroc

**Mohamed Binkkour**
ISIAM, Av. Hassan I, B.P 805, Agadir, Maroc
mohamed.binkkour@yahoo.fr
Institut Supérieur d'Informatique Appliquée et de Management ISIAM, Agadir, Maroc
Centre de Recherche Appliquée en Management (CRAM) Agadir, Maroc

**Abstract :**
This paper aims to present the different aspects and characteristics of strategic and operational information and propose a categorization pattern allowing to consider an information as strategic or operational. This categorization is to be used in the two decision making processes to assist its mining, and usage by the two related decision support systems. This is conducted trough the results of an investigative study of information used as basis for strategic decisions inside three different companies.

**Résumé :**
Cet article a pour objectif de mettre en évidence les caractéristiques des informations stratégiques et opérationnelles et de proposer une grille de classification permettant de considérer une information comme stratégique ou opérationnelle. Cette classification d'information est établie dans le but de pouvoir l'exploiter dans les deux types de processus de prise de décision et faciliter son extraction et usage par les deux systèmes de support de décision y afférents. Ceci est réalisé à travers les résultats d'une étude exploratoire des informations bases de décisions stratégiques au sein de trois différentes compagnies.

**Mots-clefs :**
Informations stratégiques, Informations opérationnelles, DSS, ESS, SIS, EIS, data mining, data warehouse, text mining

# Strategic and Operational information
# Support of decision making processes and systems

## INTRODUCTION

Nowadays, the level of success for an organisation is based on its capacity to leverage a considerable amount of data and translate it into successful decisions.

With the boom of information, the multiplication of operational information systems (ERP, SCM, CRM, DWH, …) [DIRK2004] and the growing in complexity systems of data mining and data warehousing and competitive intelligence, it is no more a simple choice to make to generate information capable to stand a foundation for optimal strategic decisions.

In this paper we will try to explore the strategic and operational information in their relation with the competitive intelligence in terms of decision making, and examine criterias of differentiation between strategic and operational information.

## I- FROM DATA TO DECISION

In organizations, a distinction is quite explicit between strategic decisions and executive decisions. In a general rule, a strategic decision is the one that is likely to affect long term direction of an organization to give some internal or environmental advantage towards its objectives [JOHN1997]. While Executive or operational decisions are more likely to affect short term management of the organization to run periodic business process activities to reach short term goals.

These two types of decisions lie on two support systems : Decision Support System (DSS) –which is part of the Strategic Information System- and Executive Support System (ESS) –which is part of the Executive Information System- that provide management with data and information enabling them to make the one or the other type of decisions. *(See **Fig.1**)*

According to Wiseman, Galliers & Somogyi, the Strategic Information System (SIS) is a system to manage information and assist in strategic decision making [CHAR1987] that aims to enhance competitiveness of the organization in its environment trough the application of IT to business process. While Executive Information System (EIS) is designed to improve efficiency and effectiveness by automating back office data processing functions and improving information flows and transfers [ELIZ1987].

Thus, we dispose of two categories of information :
Strategic Information [ERKK1989] and Operational (Executive) Information.

Although the distinction between strategic and operational decisions, support systems and Information Systems is quite clear, it is not as clear in the case of Strategic and Operational Information due to the little mention of it in library and information science literature [ERKK1989].

Both types of information –with some prevalence of strategic information- are collected, processed and diffused according to the competitive intelligence process to generate new information that can help in decision making [PRES2001],[MIRE2000], [BARN1986].

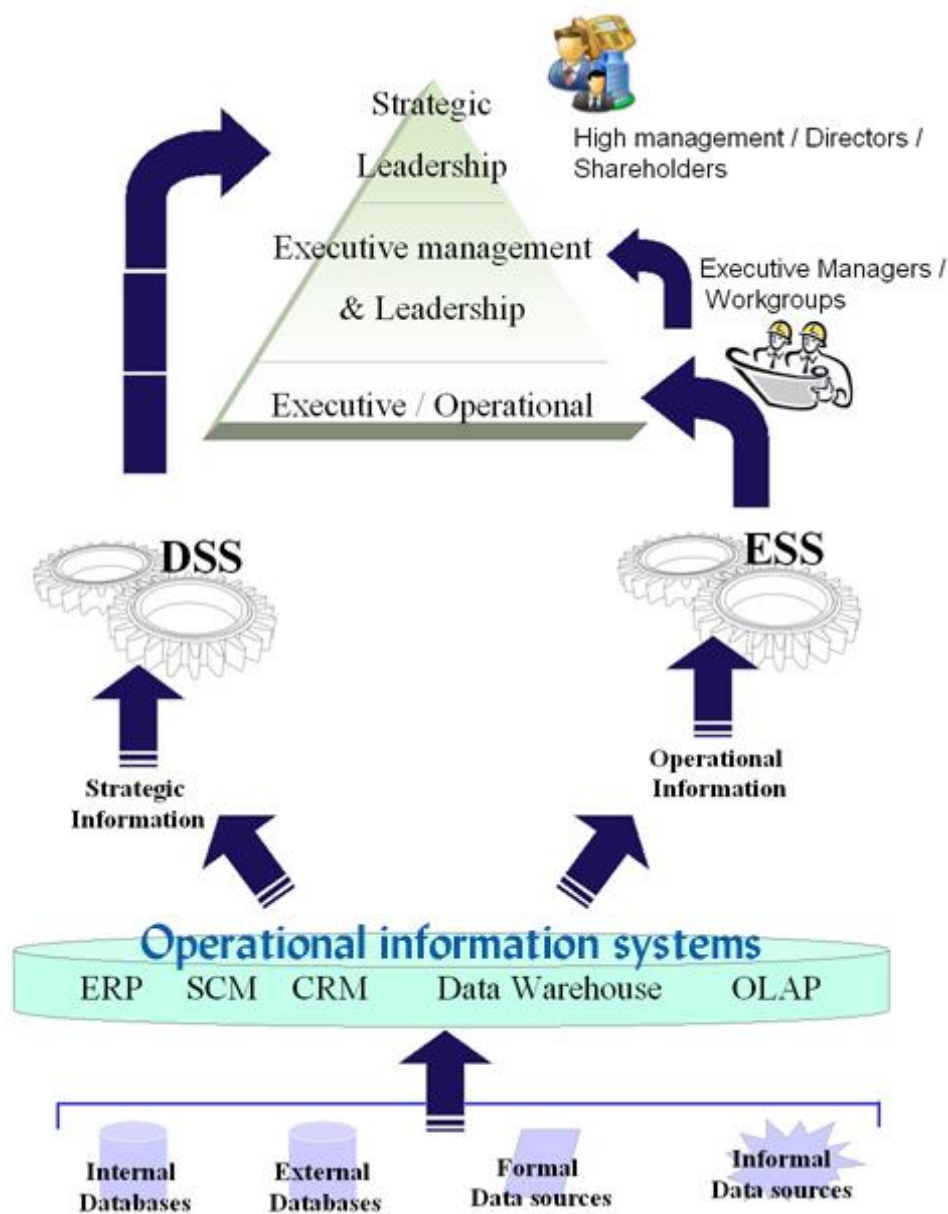

**Fig.1** Process of decision making

In databases and data warehouses, data is stored mainly unprocessed in its raw state. With the advance in I.T and data operational systems, data is regularly processed to produce strategic information. However, the distinction between information designed to be used in strategic decision making and the one to be used in operational decision making is still ambiguous which makes the task of extracting and processing of information to support the two different types of decisions difficult and subjective.

Therefore two questions need to be asked:
**How to build and organize data mining, text mining, data warehousing and competitive intelligence models and systems to fit both strategic and operational information needs in term of decision making?**
**How to build models to translate Competitive Intelligence procedures, actions and systems into assisted strategic decision making models and systems?**

To propose answers to these two questions we need first to differentiate between information called strategic and the one that is operational or executive, based on the characteristics of each. This is conducted trough the results of an investigative study of information used as basis for strategic decisions inside three different companies.

## II- CATEGORIZING DATA INTO STRATEGIC AND OPERATIONAL

For an information to be valuable in a decision system, it has to present some main qualities [ROBE2004], [ILM2003], [CHCK2000] depending on its use.
Information has to be :
Accurate, complete, correct, reliable, available, cost-beneficial, user-targeted, relevant, authoritative, timely and easy to use.
Moreover, strategic and executive data intended to support strategic and executive decisions have additional characteristics, related to their source, usage and presentation.

**We need to explore these characteristics in detail and try to categorize data -in a data warehouse for example- so to tell if such data is oriented strategic or executive in an organisation .**
**Mainly, we differentiate between strategic and operational data according to the type of decisions that rely on it. So a strategic data or information** [ALLEN1994] **is the one used in the strategic decision making process, while an operational (executive) data is used in the operational decision making process**[ERKK1989]**.**

Criterias of distinction are

### II.1 Frequency of production
The strategic data is generally produced on demand, on an 'ad hoc' basis as a response to high management request. In the other hand, operational data is produced routinely on a regular basis to illustrate the continuous activity of the organization.

### II.2 Relevance of the information
As stated earlier, Strategic data or information is relevant to the strategic decision making that means long term. While executive data and information is relevant to the short and medium term.

### II.3 Quantitative and qualitative aspects of data
Long term decisions are based on both quantitative and qualitative information, while executive decisions are drawn mainly from quantitative information.
As an illustration we could take the level of innovation for a given product for example which is very difficult to evaluate or quantify mathematically.

### II.4 Source and generation of data
Strategic data is often derived from both internal and external sources, while executive data is primarily generated internally, but may have limited external components. The external extent of strategic decision –such as market share evolution strategy for a company- requires an extensive use of external data and information. Strategic decision are large scope decisions requiring to be based on trustful internally and externally generated information. Operational decisions, in the opposite, concern in general, work groups or departments inside the organization and are less influenced by the external environment and data.

**II.5 Scope of data concern**
Aligned with the strategic management of the organization, strategic information is concerned with the whole organization and its long term objectives that have as a main characteristic being global and integrated. [JOHN1997] The major difficulty is to define and confirm the meaning of "whole organization" which is usually subjective and dependent of the size, sectors of activity and type of leadership in each organization.

**II.6 Access and use frequencies**
In an organization, some executive data might be used daily or on a regular bases depending on the business cycle that might range from one day to 3 months. Strategic data, in the other hand, is less frequently used. Comparing for example the use of a 'bank account statement' with a 'dividend per share'.

**II.7 Update frequency**
Operational data is updated on a regular basis (usually based on the business cycle of the department or work group). This is operated at low levels of hierarchy, such as finance, sales etc. While strategic decision supporting data is promptly updated, on demand and not necessarily on a regular bases, such as data utilized for general assembly, and shareholders meetings, and decisions.

**II.8 Level of aggregation**
For a company the percentage of bankruptcy within its raw materials suppliers –for example- is an important aggregated information that might draw the future trends, and strategies. While a single bankruptcy case is less to be considered in such kind of decisions. In a word, strategic information and data is usually presented in aggregated form in opposite with operational information.

**II.9 Level of detail**
As mentioned above, in databases and data warehouses data is mainly stored in its raw state, with a high level of detail, such as a row in the customer table which might be used on a daily bases to make operational decisions and conduct appropriate business actions. Despite the fact that this level of detail preserves the value of data, it is less appropriate to be considered as a basis when dealing with strategic decision making process.

**II.10 Data subject (single raw content)**
In relational databases, a single row in a table 'A' has usually one or more associations with rows in other tables within the database. On a regular basis, this fact helps avoid redundancy and optimise access and use of data, allowing a considerable gain in time and performance. While Strategic Data with its high level of aggregation is less apparently to be dependent of a single row content in a specific table within the database.

**II.11 Scope of data production and generation**
**(Produced and summarized at which hierarchical level?)**
In general, strategic information is produced and summarized at a high level of the hierarchy in the organization, while executive data is produced within the range of its use i.e. operational level of the hierarchy. *(See **Fig.2**)*

If data production is supervised by Strategic level managers, even though, produced at the managerial level of the organisation, we might categorize it as strategic depending on the supervising operated during its production. Otherwise it usually is related to executive decision making i.e. operational.

We can present as example the annual financial statements edited under supervising of the higher management in opposition with weekly sales report produced, summarized and supervised by middle management, or the daily storage capacity report generated at the executive level.

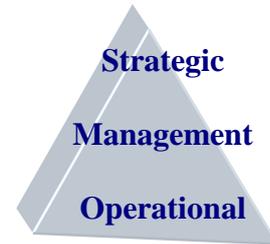

**Fig. 2**- Hierarchical levels

**II.12 Use of dating**
The fact of using a date in some data implies its punctual aspect. Generally dated data is used at the executive level of the organization rather than the strategic one. Dated information -such as dated invoices or transactions- are apparently not appropriate for the long term decision making process.

**II.13 Use in statistical analysis and regression**
Both strategic and executive levels of the organization rely on statistics to make and justify their decision making. However, the use of statistical regressions and correlation is basically used for projections and long term forecasting.

**II.14 Formal / Informal presentation of data**
Formal presentation of data points to data organized and presented according to defined and standardized rules in tables, databases, data warehouses etc. While informal presentation of data luck this aspect of organization even tough respecting other conventions and rules.

Necessarily operational data rely on formally presented data, as data is in its biggest part quantitative. However, we notice that in the case of strategic decision making process, even non formally presented data is used as long as it respects the characteristics of a valuable information –mentioned earlier-, and concerns the long term. Take the laws outlines against tobacco for example, in their relation with the tobacco giant companies decisions.
.

**II.15 Source operational information system**
We mean by Operational Information Systems, tools and techniques used to extract, manipulate, integrate and present varied data from one or more data sources generating final data output designed to serve as a basis for decision making.
Various tools belong to this group of I.S, such as ERP, SCM, CRM, Data warehouses …
Some of these tools are more adequate for strategic decision making than others. A more in-depth comparative study needs to be conducted to propose a better classification in term of decision making support capability of each category.

**II.16 Data access, availability and destination**
The question is to determine if data is general access, restricted access or exclusive access. In general, strategic data is defined as top secret data with exclusive access. It also might be characterized by restricted access as many departments may have access privileges to it.
Concerning executive data, it ranges from exclusive access data to general access data according to its use and sensibility to security issues. This aspect of data access availability and destination is very noticeable in the cases of new product launches, inventions, innovations and so on.

### II.17 Purpose of data
In an Information System, data might fill analytical, historical or informational need or a combination of these three types. Based on the purpose of data, we remark that strategic decision systems lie widely on historical and analytical data, while operational decision systems use essentially informational in combination with analytical data.

### II.18 Abstraction level
In contrast with operational data which is very specific and well defined, strategic information might be presented in an abstract way. Many strategic decisions lie on combinations of personal judgments, general ideas, subjective experiences and immeasurable appreciative judgments.
As example, the Goodwill for a specific company is very abstract concept that is used generally to make long term investment decisions in Financial Markets.

### II.19 Vertical / Horizontal data
Executive data is concerned with activities or departments. In other terms, concerned with the vertical aspect of the organization. At the same time, strategic data is cross activities and cross departments synchronously generated to attend a specific objective. Strategic data reflects the horizontal aspect of the organization.
Compare, for this situation, decisions related to customers satisfaction which is no more a concern of the sales department alone to the auditing of a specific customer account.

### II.20 Forecasting use of information
Strategic information is produced in an optic of forecasting, knowing that this kind of information is quite uncertain, as the future cannot be predicted accurately. To this is added the tacit characteristic of knowledge which is usually a basis for decision making in a strategic perspective. [NONA1998].

### II.21 Long term tracking data
The general characteristic of strategic data is that it keeps track of long term evolution of facts and actions. Basically, this tracking is translated in statistics trough the use of years as the basic timeline. We might, for instance, compare logistic decisions based on the daily weather forecast to long term investment decisions based on the climate change evolution data.

### II.22 Default processing mechanism of data
In organizations, operational data has known precise data circuits usually in form of procedures, laws and directives. While procedures regarding strategic data are loose and flexible depending on the type of decision to be made.

### II.23 Investment data
An investment decision is mostly a long term decision. Most data issued from investment related functions and departments serve as background for log term decision making.

**CONCLUSION**

Finally, by exploring the differences between strategic and operational information trough this study, we might build a first comparative scale to differentiate between the two categories of information. Knowing that on these kind of information we base decisions that might cost the organization very expensive price in term of competitiveness and survival, we feel the importance of having a clear image and a concise understanding of which information is strategic so to use it for strategic decision making and which is operational we can base our executive decisions upon. The next step is to design models allowing to rise these characteristics in the divers data automated systems so as to auto-differentiate between the two categories, and much easily provide the appropriate decision system with the appropriate type of data.